# Photonic integrated circuit polarizers based on 2D materials

*David Moss*

Optical Sciences Centre Swinburne University of Technology Hawthorn, VIC 3122, Australia

Email: dmoss@swin.edu.au

**Abstract**

Optical polarizers are essential components for the selection and manipulation of light polarization states in optical systems. Over the past decade, the rapid advancement of photonic technologies and devices has led to the development of a range of novel optical polarizers, opening avenues for many breakthroughs and expanding applications across diverse fields. Particularly, two-dimensional (2D) materials, known for their atomic thin film structures and unique optical properties, have become attractive for implementing optical polarizers with high performance and new features that were not achievable before. This paper reviews recent progress in 2D-material-based optical polarizers. First, an overview of key properties of various 2D materials for realizing optical polarizers is provided. Next, the state-of-the-art optical polarizers based on 2D materials, which are categorized into spatial-light devices, fiber devices, and integrated waveguide devices, are reviewed and compared. Finally, we discuss the current challenges of this field as well as the exciting opportunities for future technological advances.

**Introduction**

The control of light polarization plays a critical role in advanced optical technologies, underpinning the functionality and efficiency of modern optical systems [1, 2]. Optical polarizers, which selectively transmit light with a specific polarization orientation and block light with the orthogonal polarization, are crucial components to realize light polarization control for a diverse range of applications as illustrated in **Fig. 1**. The applications in **Fig. 1** are divided into four main categories, including optical sensing, imaging and display, optical analysis, and optical communications. For optical sensing applications such as navigation, astronomical detection, and atmosphere detection [3, 4], optical polarizers enhance the sensitivity and accuracy by selecting the desired polarization state of incident light. The use of optical polarizers for imaging and display includes applications such as glasses, liquid crystal display, and photography [5, 6], where optical polarizers help reduce glare, enhance contrast, and improve the color reproduction in displays, contributing significantly to image quality and visibility. For optical analysis such as spectroscopy, biomedical measurements, and materials analysis [7, 8], optical polarizers have been employed in various analytical techniques to extract valuable information from measured optical signals. In optical communication systems, optical polarizers have been widely used for implementing laser systems, reducing polarization-mode dispersion, and realizing polarization-division multiplexing [9, 10], thereby maintaining signal integrity and achieving efficient data transmission.

The rapid progress of the photonics industry urgently requires high-performance optical polarizers on different device platforms and at extended wavelengths, which has posed significant challenges for conventional optical polarizers based on bulk materials [11]. Since the first experimental isolation of graphene in 2004 [12], the family of two-dimensional (2D) materials has been growing rapidly and become a hotbed of scientific inquiry. Many 2D materials, contrasted with their bulk counterparts, are renowned for their extraordinary material properties. Amongst them, the attractive optical properties of 2D materials, such as layer-dependent optical bandgaps, strong quantum confinement effects, tunable light emissions, high optical nonlinearity, and significant material anisotropy [13-16], have been the cornerstone of extensive research interests for their utilization in implementing functional optical devices.

In recent years, owing to their highly anisotropic light absorption and broad response bandwidths, the use of 2D materials to manipulate light polarization states has come into the spotlight [17-22]. This opens new avenues for the implementation of optical polarizers, enabling high device performance and new features previously unattainable with conventional bulk materials. In addition, the fast progress in fabrication technologies enables the incorporation of 2D materials into different device platforms. This facilitates the development of a series of advanced optical polarizers based on spatial light devices, fiber devices, and integrated waveguide devices, significantly enhancing the range of application scenarios.

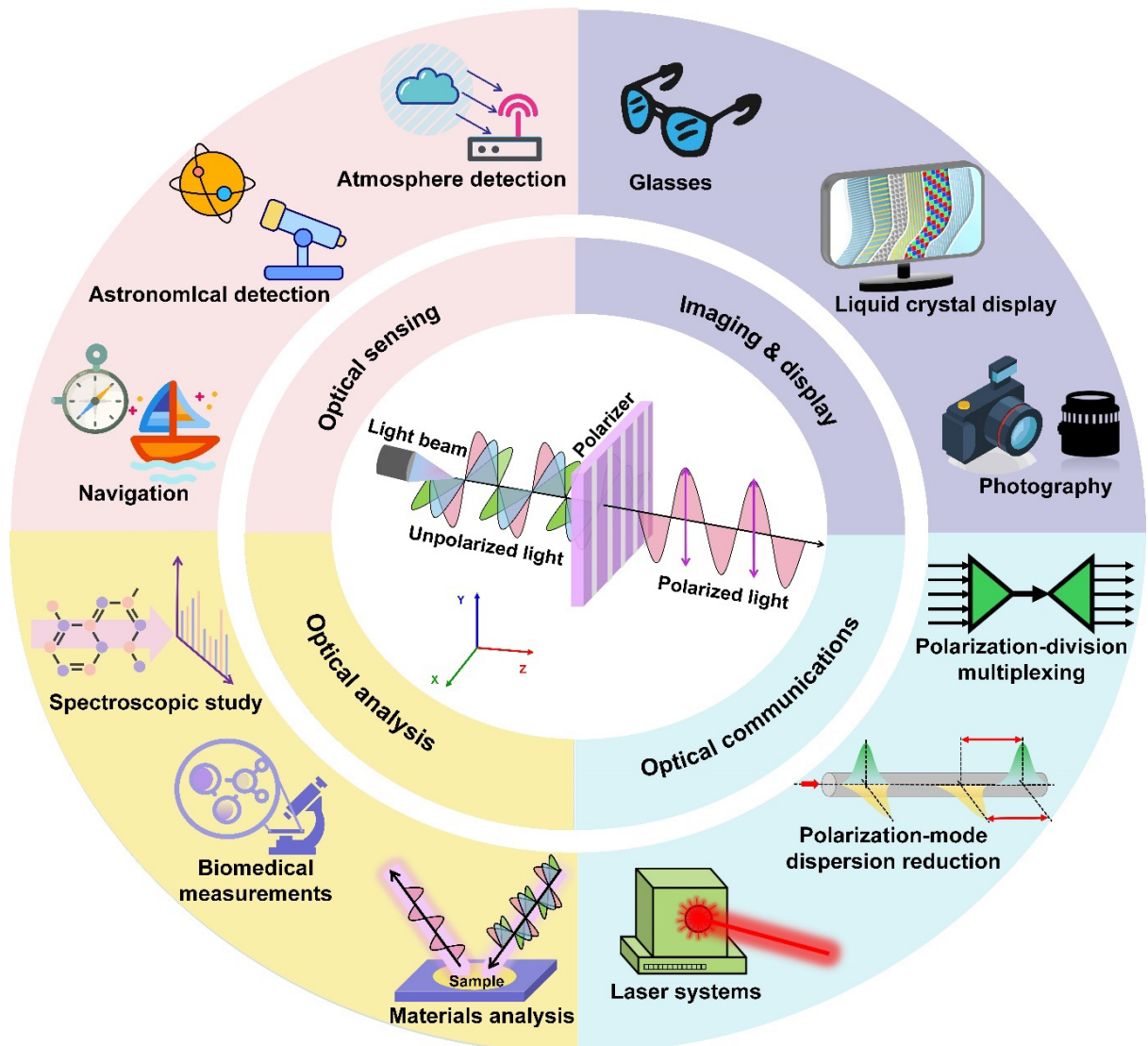

Fig. 1. Typical applications of optical polarizers.

Here, we provide a systematic review of 2D-material-based optical polarizers, highlighting the significant progress achieved in this field and offering performance analysis of devices in different platforms. This paper is organized as follows: First, the distinctive material properties of 2D materials are introduced within the context of their applications to optical polarizers. Subsequently, different types of 2D-material-based optical polarizers are reviewed and compared, which are classified into spatial-light devices, fiber devices, and integrated waveguide devices. Finally, we discuss the present challenges faced by this field, alongside the promising prospects that pave the way for future technological breakthroughs.

## Properties of 2D materials for optical polarizers

2D materials with atomically thin film thickness exhibit anisotropic absorption for light in different polarization states, which forms the basis for their applications to optical polarizers [13]. As illustrated in **Fig. 2a**, 2D materials typically show higher absorption for in-plane polarized light (*i.e.*, with its optical field *E* parallel to the surface of 2D material) that enables stronger light-matter interaction, as compared to light in the out-of-plane polarization (*i.e.*, with its optical field *E* perpendicular to the surface of 2D material). For 2D-material-based optical polarizers, the difference in absorption for light with different polarization states determines the polarization extinction ratio (PER, defined as the ratio of transmitted power for the desired polarization over the undesired one), and the capability to maintain such anisotropy in broad optical bandwidths influences their operational bandwidths (OBW). In addition, some other properties of 2D materials play important roles in enhancing device functionality. For instance, low optical absorption of 2D materials in the high-transmittance polarization state is desired for minimizing the additional insertion loss (IL) induced by them, and the ability to alter material properties through gate voltage or light power enables the realization of devices with dynamic tunability. In this section, we provide an overview of distinctive properties of typical 2D materials illustrated in **Fig. 2b**, highlighting those that are particularly relevant and advantageous for implementing high performance optical polarizers.

### *Graphene*

Graphene consists of single-layered carbon atoms arranged in a hexagonal lattice, where the linearly dispersive conduction and valence bands meet at the Dirac point of the Brillouin zone, resulting in a gapless and semi-metallic band structure [23, 24]. This unique structure endows graphene with the capability to interact with light waves in different directions and across ultrabroad optical bandwidths. Despite its ultra-low film thickness of ~0.35 nm, monolayer graphene can absorb ~2.3% of light power incident perpendicularly [25]. Such absorption could introduce additional insertion loss for graphene-based optical polarizers. Due to intraband transition and its property related to surface plasmon resonance, graphene can also strongly interact with terahertz (THz) light waves, making it promising for optical polarizers operating at this spectral range [26, 27]. In addition, the zero bandgap in graphene can be further tuned by applying external electric fields or through chemical doping, enabling the flexible changing of its properties to enhance the device performance [28-30].

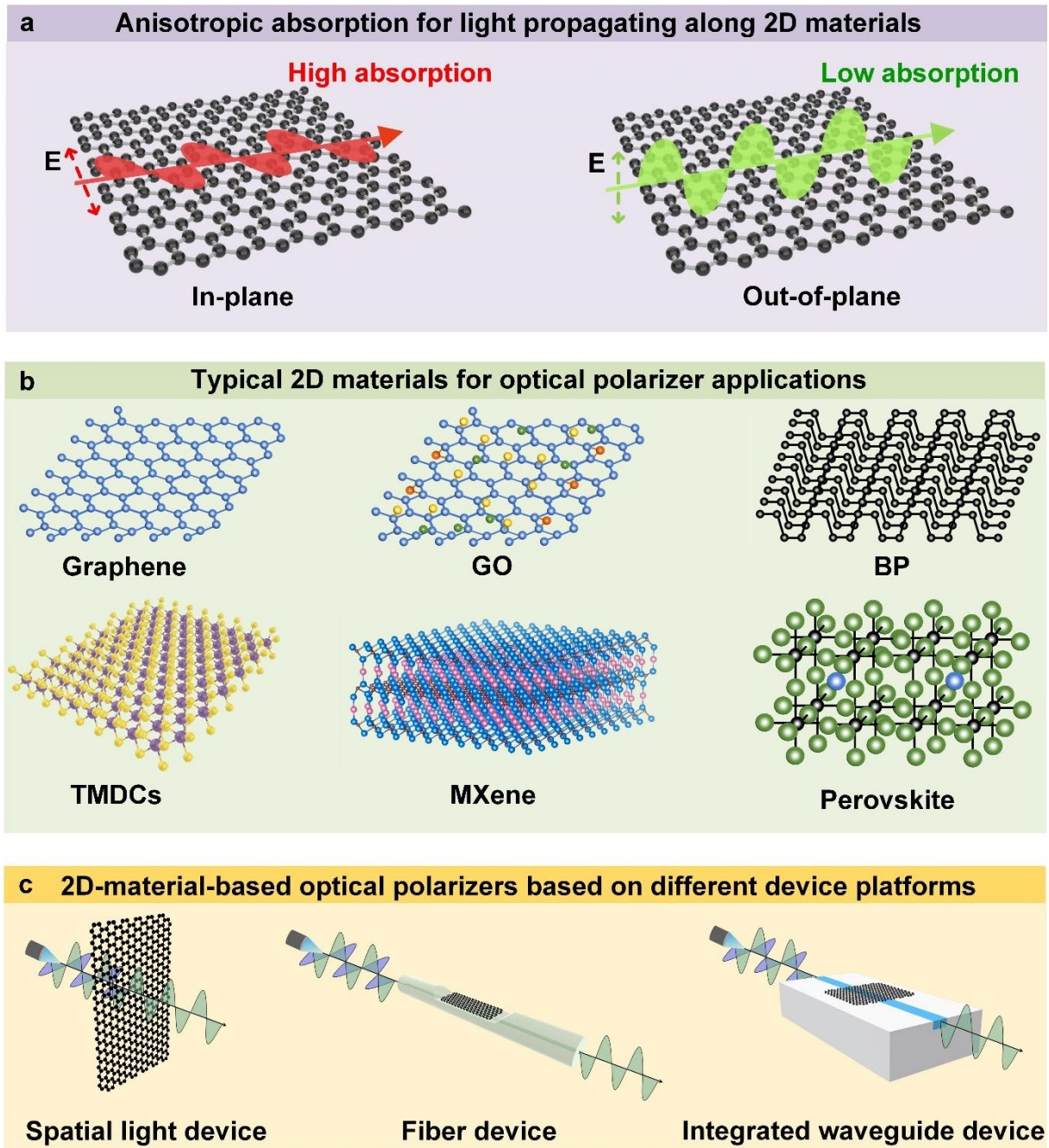

**Fig. 2.** (a) Anisotropic absorption for light propagating along 2D materials. (b) Typical 2D materials

for optical polarizer applications. (c) 2D-material-based optical polarizers based on different device platforms. GO: graphene oxide. TMDCs: transition metal dichalcogenides. BP: black phosphorus.

## *Graphene oxide*

Graphene oxide (GO), as a common derivative of graphene, consists of a basal plane of carbon atoms decorated with a series of oxygen-containing functional groups (OCFGs) such as hydroxyl, carboxyl and carbonyl groups [31]. Due to the existence of the OCFGs, a monolayer GO film, with a typical thickness of ~1 nm, has a higher thickness than monolayer graphene [32]. The OCFGs also result in a highly heterogenous atomic structure for GO, which allows it to exhibit strong anisotropic absorption for light with in-plane and out-of-plane polarizations [33]. It is also observed that such anisotropic light absorption diminishes progressively with the augmentation of film thickness [34]. The large optical bandgap of GO, typically ranging from ~2.1 to ~3.6 eV, can be altered through a variety of reduction and doping treatments [35], allowing for the engineering of optical absorption across various spectral bands. The large bandgap of GO also results in much lower optical absorption as compared to graphene that has a zero bandgap, particularly at infrared wavelengths [36, 37]. These material properties of GO, combined with its facile solution-based synthesis processes and the ability to be directly grown on diverse dielectric substrates, contribute to the attractiveness of GO for the implementation of optical polarizers [38-40].

## *Transition metal dichalcogenides*

Transition metal dichalcogenides (TMDCs) are characterized by the general formula $MX_2$, where 'M' stands for a transition metal and 'X' for a chalcogen element. TMDC layers, such as molybdenum disulfide ($MoS_2$), tungsten disulfide ($WS_2$), molybdenum diselenide ($MoSe_2$), and tungsten diselenide ($WSe_2$), have a typical sandwich-like configuration, where a plane of metal atoms is flanked by two hexagonal planes of chalcogen atoms. This distinct structure allows TMDCs to exhibit varied bandgap (*e.g.*, from ~1 to ~2.5 eV [41, 42]) as well as indirect-to-direct bandgap transition when the material thickness decreases from multilayer to monolayer (*e.g.*, $MoS_2$ with a typical thickness of ~0.72 nm) [43]. TMDCs also show significant anisotropy in light absorption (>10% per layer) that covers from the visible to the infrared wavelength regions [44, 45]. In addition, monolayer TMDCs exhibit unique valley-dependent properties, where light with a particular polarization state can excite electron transitions in specific valleys [41, 46]. For example, it is observed that right-handed and left-handed circularly polarized light can effectively excite electrons at the *K* and *K'* points (*i.e.*, the two non-equivalent corners of the Brillouin zone in a hexagonal lattice system), respectively, thereby allowing for sensitive control over the polarization states of light in these materials [47]. Recently, noble metal-based TMDCs including rhenium disulfide ($ReS_2$), palladium diselenide ($PdSe_2$), platinum diselenide ($PtSe_2$), and palladium ditelluride ($PdTe_2$) have also be utilized for polarization-sensitive devices due to their exceptional ability for anisotropic light absorption [48-50].

## *Black phosphorus*

Black phosphorus (BP), commonly referred to as black phosphorene in its monolayer form, has a layered atomic structure with phosphorus atoms arranged in a puckered honeycomb lattice. In contrast to in-plane centrosymmetry observed for graphene and TMDCs, the distinctive puckered configuration of BP breaks the in-plane centrosymmetry, which gives rise to anisotropic in-plane optical conductivity [51-53]. As a result, the optical absorption in BP exhibits anisotropy for in-plane light waves that travel in different directions, and it is influenced by several factors such as film thickness and doping levels [54, 55]. Besides, BP features a layer-dependent bandgap, ranging from ~0.3 eV in its bulk form to ~2.0 eV for a monolayer film with a typical thickness of ~0.50 nm, effectively covering the visible to near-infrared wavelengths [53, 56, 57]. This versatility enables the design of optical polarizers targeted for specific wavelength bands. Due to chemical degradation caused by the oxygen and moisture in ambient conditions, BP experiences instability upon exposure to air, posing potential challenges for its practical implementation [58]. To address such limitation, several methods have been employed, including organic functionalization, inorganic coatings, and synthesis through liquid phase exfoliation [59, 60].

## *MXene*

MXenes are a novel class of 2D materials composed of transition metal carbides, nitrides, or carbonitrides. The general formula of MXenes is $M_{n+1}X_nT_x$, where *M* represents a transition metal, *X* stands for C and/or N element, *T* denotes surface terminations such as -OH, =O, and

-F, and $n$ = 1, 2, or 3 [61]. These surface terminations enhance the water affinity of MXenes, making them easily dispersible in aqueous and other polar solvents, leading to stable colloidal solutions. The hydrophilic nature of MXenes facilitates their deposition on a variety of substrates, showing the potential for simple device fabrication [62, 63]. Depending on the composition of chemical elements, the typical thickness of monolayer MXene films varies from ~1 to ~3 nm [64]. With their unique metallic conductivity attributed to the free electrons in the transition metal carbide or nitride backbone, MXenes present a compelling electronic structure. This, combined with their surface terminations, leads to strong anisotropy in optical absorption covering broad bandwidths from the visible to the mid-infrared (MIR) regions [65]. Similar to the intraband transition in graphene, the strong optical absorption of MXenes in the THz range makes them appealing for advanced THz optical polarizers [66]. In addition, the properties of MXenes are enriched by the ability to adjust their surface functionalization, which allows for the modulation of their bandgaps within a wide range spanning from ~0.05 to ~2.87 eV [67, 68].

### *Perovskites*

2D perovskites are characterized by layered structures, comprising single (or multiple) inorganic sheets sandwiched between organic spacers held together by Coulombic forces. They have a general formula $R_2A_{n-1}B_nX_{3n+1}$, where $R$ is an additional bulky organic cation such as aliphatic or aromatic that functions as a spacer between the inorganic sheets, $n$ defines the number of inorganic layers held together and determines the thickness of the perovskite film, $A$ is an organic cation, $B$ is a metal cation, and $X$ represents a halide ($Cl^-$, $Br^-$ or $I^-$) [69]. Due to the existence of interleaved inorganic and organic sheets, perovskite materials possess a tunable bandgap that can be modified by changing their composition. For example, the bandgap of $PEA_2MA_{n-1}Pb_nI_{3n+1}$ decreases from ~2.36 eV ($n$ = 1) to ~1.94 eV ($n = \infty$) as the number of inorganic layers increases [69, 70]. Due to their high photovoltaic efficiency and strong light absorption across the solar spectrum, perovskites have been extensively utilized in solar cell applications [71, 72]. In addition, the intrinsic non-centrosymmetry of the crystal structures endow perovskites with strong optical anisotropy for the detection or emission of polarized light [73, 74], which forms their applications to polarization selective devices.

Moreover, perovskite materials demonstrate environmental stability since the organic components can provide hydrophobicity and mechanical flexibility, while the inorganic parts offer structural robustness and chemical resistance [75].

*Van der Waals heterostructures*

Exploring beyond the use of a single material, the assembly of different 2D materials into Van der Waals (vdW) heterostructures introduces many new features and possibilities for photonic and electronic devices, and these also include optical polarizers. Compared with conventional bulk heterostructures that suffer from lattice mismatch and defects at the interface when combining materials with different lattice constants, vdW heterostructures can address these challenges due to the inherent propensity of 2D materials to adhered through van der Waals (vdW) force [76, 77]. Furthermore, the interface can be atomically sharp, and the thickness can be as thin as a few atomic layers. By combining the distinct properties of various 2D materials such as anisotropic conductivity, direct bandgaps, and strong light-matter interactions, the vdW heterostructures can overcome the drawbacks of a specific 2D material. For example, 2D materials such as black phosphorus are prone to environmental degradation or possess inadequate mechanical properties. By integrating them into vdW heterostructures, the stability and durability of the composite structure can be significantly improved, as the more resilient materials afford protection to those that are more susceptible. The vdW heterostructures also enable more efficient manipulation of light polarization states. For example, an optical polarizer based on graphene/$MoS_2$ heterostructure exhibited ~1.75 times enhancement of light absorption than a comparable graphene-based optical polarizer [78]. In addition, the flexibility in designing the vdW heterostructures enables customization for the optical responses and operational wavelength ranges. This allows for tailoring of the polarizer performance parameters and the broadening of their potential applications.

## 2D-material-based optical polarizers in different device platforms

The dangling-bond free surfaces of 2D materials allow them to be readily incorporated into different device platforms without stress or restrictions due to lattice mismatch. To implement optical polarizers based on 2D materials, various devices based on different platforms have been proposed and demonstrated [17, 20, 21], each presenting unique advantages and constraints. In this section, we review and compare the performance of these optical polarizers, which are categorized into spatial-light devices, fiber devices, and integrated waveguide devices according to the device platforms illustrated in **Fig. 2c**.

## *Spatial-light devices*

Spatial-light optical polarizers allow for the selection of light polarization states in free-space environments. They are featured by high polarization selectivity across a specific wavelength range and robust environmental stability. Spatial-light optical polarizers can be realized based on various schemes with different device configurations. For instance, wire-grid polarizers utilize finely spaced metallic wires to selectively absorb light in one polarization while allowing the transmission of the other [79]. For sheet polarizers, polarization selection is achieved by absorbing one polarization state more efficiently than the other [80]. In prism polarizers, birefringent materials are employed to refract and separate light in different polarization states [81]. Brewster-Angle polarizers operate by reflecting light at a specific angle to isolate a particular polarization state [82]. As compared to conventional spatial-light optical polarizers based on bulk materials, the incorporation of 2D materials in spatial-light optical polarizers could yield broader OBWs and flexible polarization selection through thermal or electrical tuning methods. In addition, 2D materials have a high mechanical flexibility to maintain their functionality when subject to bending, twisting, or compression. In **Fig. 3**, we summarize typical spatial-light optical polarizers incorporating 2D materials.

As mentioned in the previous section, graphene can effectively interact with light at THz wavelengths. This has motivated many studies on THz graphene optical polarizers. In 2012, Bludov *et al*. first proposed a THz graphene spatial-light optical polarizer and theoretically analyzed its performance. In such device, a graphene layer was sandwiched between two dielectric media and a prism was placed on top of them **(Fig. 3a)** [83]. Polarization selection was realized by leveraging the resonant coupling between transverse magnetic (TM)-polarized light waves and graphene surface plasmons, where TM-polarized waves were absorbed at resonance, and transverse electric (TE)-polarized waves were reflected. By adjusting the voltage applied to graphene, it was possible to modulate the degree of polarization change and achieve 100% reflection of the TE-polarized waves. This could also enable the transition between circular and linear polarizations.

Metasurfaces consisting of 2D arrays of elements have been widely used for manipulating anisotropic propagation of light waves [84-86]. By incorporating graphene layers into metallic metasurfaces [20, 26] or utilizing purely graphene-based metasurfaces [27, 87, 88], tunable THz polarizers have been realized.

In 2017, Kim *et al.* experimentally demonstrated effective polarization manipulation of THz waves by incorporating a graphene layer into chiral metasurfaces composed of subwavelength metallic building blocks (**Fig. 3b**) [20]. In such a device, the gate voltage applied to graphene selectively adjusted the transmission of right-handed circularly polarized light without impacting the transmission of left-handed circularly polarized light, allowing for large-intensity modulation depths (> 99%) at a low gate voltage of ~2 V. In contrast to the device in **Fig. 3b** where the graphene sheet spanned the entire area of metallic metasurface, Meng *et al.* experimentally demonstrated a tunable THz spatial-light optical polarizer based on a graphene-metal metasurface with graphene wire grids positioned over a metallic metasurface (**Fig. 3c**) [26]. Metal patterns were first fabricated to deposit Ti/Au layers, followed by transferring graphene onto the metal arrays via a wet transfer process and etching to form the wire-grid structure via oxygen plasma etching. The device with interleaved graphene and metal elements enhanced the THz wave-plasmon resonant coupling, enabling tunable PER from ~3.7 to ~10.3 dB by adjusting the gate voltage.

Although the light-graphene interaction can be enhanced by using metallic metastructures, they also face limitations, such as complicated fabrications and high costs. To address this issue, some THz optical polarizers based on pure graphene metasurfaces were investigated. A THz optical polarizer was proposed by placing an array of rectangular or L-shaped graphene patches on top of a dielectric layer to form metasurfaces (**Fig. 3d**) [87]. Owing to the strong plasmonic behavior of graphene, the rectangular graphene patches facilitated the conversion of linearly polarized waves into circular or elliptical polarization during transmission, while the L-shaped graphene patches enabled the conversion of linearly polarized waves into their cross-polarized counterparts during reflection. In addition, a tunable THz optical polarizer with adjustable operating frequency within 1.81-3.81 THz was theoretically investigated (**Fig. 3e**) [27], where a graphene ribbon array on silicon substrate passed TM-polarized light and blocked the TE polarization due to the surface plasmon resonance of graphene. By employing a double-stage graphene layered structure in such a device, an optimized PER up to ~30 dB was achieved. Similarly, a THz graphene wire-grid optical polarizer was proposed by utilizing the surface plasmon resonance in graphene ribbons to select TM-polarized light [88]. Simulation results showed that an average PER of ~30 dB and a low IL of ~2 dB over a broad frequency range of 0.8–2.5 THz was achieved.

Apart from graphene, other 2D materials have also been employed in spatial-light optical polarizers. A mid-infrared (MIR) spatial-light optical polarizer formed by periodic GO ribbons was demonstrated (**Fig. 3f**) [89], achieving a high PER (~20 dB) as well as adjustable operational wavelengths in the MIR region. The device was designed based on the guided-mode resonance theory to allow more efficient coupling between TE-polarized light and the GO film as compared to TM-polarized light. In addition, $Ti_3C_2T_z$ MXene films with anisotropic optical conductivity were used in a wire-grid configuration to effectively manipulate the polarization of THz radiation (**Fig. 3g**) [66]. The devices were made by spinning a water-based mixture of $Ti_3C_2T_z$ nanosheets onto a photolithographically patterned THz-transparent base before soaking in acetone. Experimental results showed that the polarizer achieved a PER of ~6 dB over the spectral range from ~0.3 to ~2.0 THz.

An optical polarizer based on 2D perovskites was also demonstrated, which enabled the conversion of deep ultraviolet incident light into polarized photoluminescence of the 2D perovskite at visible wavelengths (**Fig. 3h**) [75]. The polarizer was consisted of a periodic sub-30-nm 2D $BA_2MA_2Pb_3Br_{10}$ perovskite line pattern, which was fabricated by nanoimprinting a precursor film of the 2D perovskite via spin coating on a silicon substrate with a topographically prepatterned hard PDMS mold. A thin aluminum metal layer was deposited on the perovskite nanopattern to enhance the transmission of linearly polarized light perpendicular to the line direction. The patterned 2D perovskite films with the narrow pitch size of 30 nm significantly improved the optical anisotropy, where the maximum PER was ~1.5 times higher than the pitch size of 100 nm over the spectral range of 200 – 600 nm. In addition, a BP reflective optical polarizer was demonstrated (**Fig. 3i**) based on mode interference in a multilayered system [90]. The polarizer consisted of an anisotropic BP layer atop a multilayer $SiO_2$/Si substrate, where the BP layer and the bottom substrate form a Fabry-

Perot (FP) cavity. By adjusting the thickness of the BP layer with a high optical anisotropy, a PER of ~9 dB was achieved at visible wavelengths.

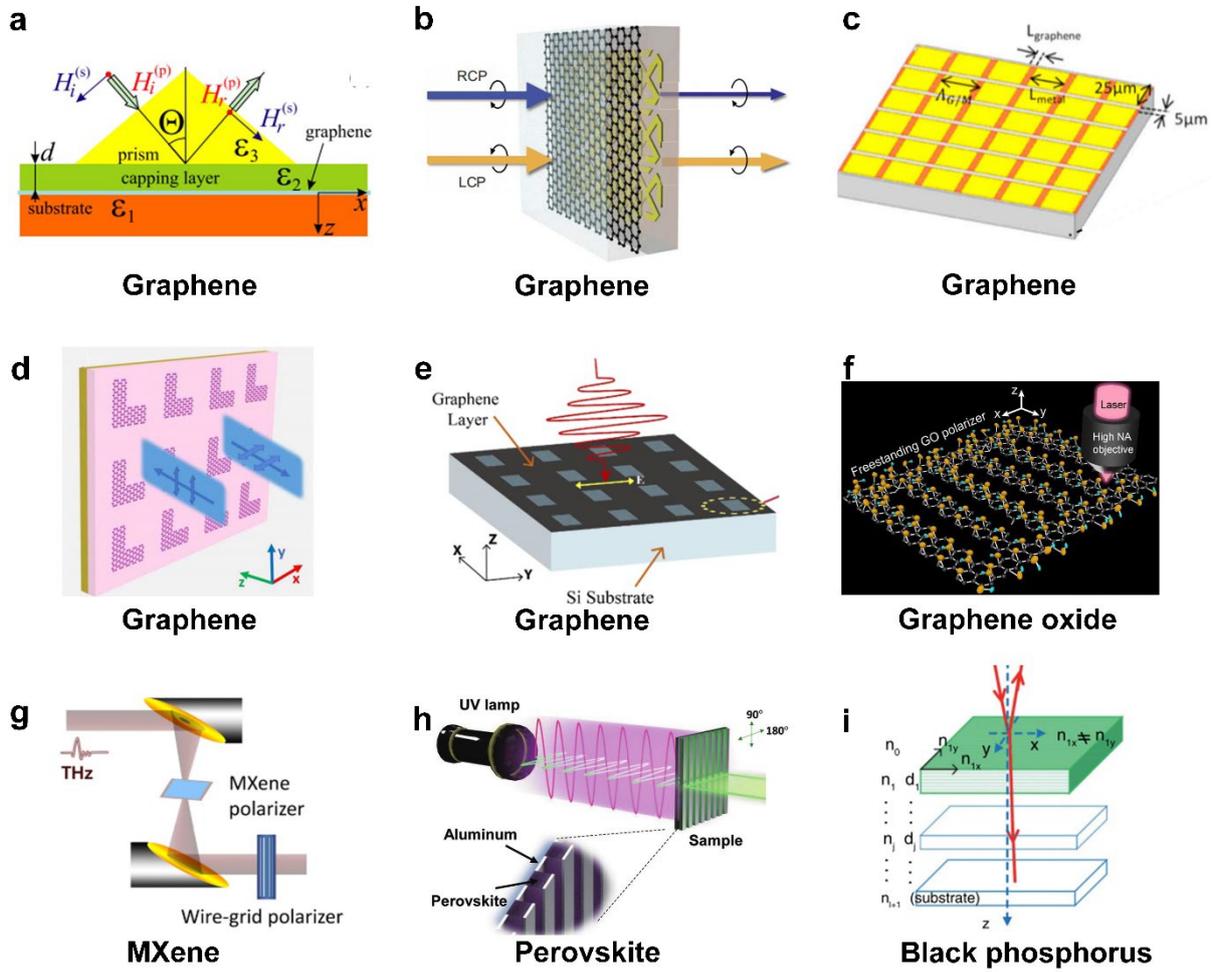

**Fig. 3. Schematic illustrations of spatial-light optical polarizers based on 2D materials. a.** A THz optical polarizer based on a graphene layer sandwiched between two dielectric media and with a prism on top of them. **b.** A tunable THz optical polarizer based on graphene incorporated into chiral metasurfaces. **c.** A tunable THz optical polarizer based on graphene-metal metasurfaces. **d.** A THz optical polarizer based on metasurfaces formed by L-shaped graphene patches. **e.** A tunable THz optical polarizer based on a graphene nanoribbon array on a silicon substrate. **f.** A mid-infrared graphene oxide (GO) optical polarizer. **g.** A THz MXene wire-grid optical polarizer. **h.** A perovskite optical polarizer working at visible wavelengths. **i.** A black phosphorus (BP) optical polarizer working at visible wavelengths. **a** is reprinted with permission from ref. [83] Copyright 2012 AIP Publishing. **b** is reprinted with permission from ref. [20] Copyright 2017, AAAS. **c** is reprinted with permission from ref. [26] Copyright 2019 OSA Publishing. **d** is reprinted with permission from ref. [87] Copyright 2016 OSA Publishing. **e** is reprinted with permission from ref. [27] Copyright 2021 OSA Publishing. **f** is reprinted with permission from ref. [89] Copyright 2021 RSC Pub. **g** is reprinted with permission from ref. [66] Copyright 2020 Wiley-VCH. **h** is reprinted with permission from ref. [75] Copyright 2023 Wiley-VCH. **i** is reprinted with permission from ref. [90] Copyright 2018 OSA Publishing.

*Fiber devices*

Fiber optical polarizers provide in-line polarization control and find direct applications in fiber-optic systems. Fiber devices are highly effective in applications like sensing and optical communications, where maintaining the integrity of the polarized light over long distances is crucial. Conventional fiber optical polarizers achieve the selection of light polarization states by incorporating birefringent crystals or metals that show polarization-selective coupling with the evanescent fiber optical field [91-93]. Due to inherent limitations of these materials (*e.g.*, polarization mode dispersion and high insertion loss [94]) and structural constraints in fiber optical polarizers, these devices face challenges in terms of fabrication, bandwidth, and tunability. In contrast, the incorporation of 2D materials with highly anisotropic light absorption over broad bands can enhance the efficiency for polarization selection and improve the OBW for fiber optical polarizers. In **Fig. 4**, we summarize typical fiber optical polarizers incorporating 2D materials.

In 2011, Bao *et al.* first demonstrate a graphene optical polarizer in a single-mode fiber, achieving a high PER of ~27 dB and an IL of ~5 dB (**Fig. 4a**) [17]. To enhance the interaction between graphene and the evanescent optical field, the fiber was polished into the core with a depth of ~1 μm. Owing to the broadband light absorption of graphene, the polarizer operated from visible to NIR wavelengths. Compared to conventional metal-clad polarizers, the graphene fiber optical polarizer also provided advantages of simpler fabrication and lower cost. Subsequently, there were many related studies aiming to improve the performance of graphene fiber optical polarizers by using various methods to enhance the light-graphene interaction. These include engineering fiber core geometry (**Fig. 4b**) [95], optimizing fiber core or graphene film position (**Figs. 4c** and **4d**) [96, 97], and introducing auxiliar materials like polymethyl methacrylate (PMMA) (**Fig. 4e**) [98] and polyvinyl butyral (PVB) (**Fig. 4f**) [99].

In addition to graphene, fiber optical polarizers incorporating other 2D materials were also investigated. By using the drop-casting method to coat GO on a non-adiabatic microfiber, a TM-pass optical polarizer was demonstrated (**Fig. 4g**) [100]. A high PER of ~37 dB at 1480 nm was achieved for a device with a ~24-μm-thick GO film at a length of ~12 mm. An optical polarizer consisting of a 2D $ReS_2$ film transferred onto a D-shaped fiber was demonstrated

(**Fig. 4h**) [101]. Due to the periodical change of linear absorption with the polarization angle, the ReS$_2$-covered fiber exhibited different responses for TE and TM polarizations, which was exploited to achieved polarization selection from the visible to MIR regions.

Recently, some fiber optical polarizers incorporating vdW heterostructures were also reported. A fiber optical polarizer consisting of MoS$_2$/graphene/Au film was demonstrated, achieving a high PER of ~19 dB (**Fig. 4i**) [78]. The polarizer was fabricated by wet transferring a graphene/MoS$_2$/PMMA hybrid film onto a side-polished fiber, followed by employing a mask to pattern interdigitated Au electrodes atop the graphene surface. In such a device, the MoS$_2$ film and the Au finger electrode significantly enhanced TM mode absorption by ~1.75 and ~24.8 times than a comparable device with only graphene, respectively. Another fiber optical polarizer based on graphene/ hexagonal boron nitride (hBN) heterostructure was also proposed (**Fig. 4j**) [102], where the graphene and hBN material were sequentially deposited on the surface of a silicon-core microfiber. Compared to the device with only graphene, more power was coupled into the graphene film to enhance the light-graphene interaction. Depending on the input wavelength and the diameter of the silicon-core, the device had the capability to function as either a TM or TE mode polarizer.

Except for using side-polished fibers in the above studies, other fiber platforms and structures were also investigated. For example, a graphene helical microfiber polarizer was designed by wrapping a microfiber on a rod coated with a graphene sheet (**Fig. 4k**) [19]. This structure enabled ultralong light-graphene interaction, yielding a high PER ~16 dB and a broad OBW from ~1200 nm to ~1650 nm. In addition, a graphene-coupled microfiber polarizer was demonstrated by attaching an optical microfiber onto a polydimethylsiloxane (PDMS) substrate coated with a graphene monolayer (**Fig. 4l**) [103]. Experimental results showed that a high PER of ~31 dB was achieved for a microfiber with an optimized diameter of ~3.9 μm.

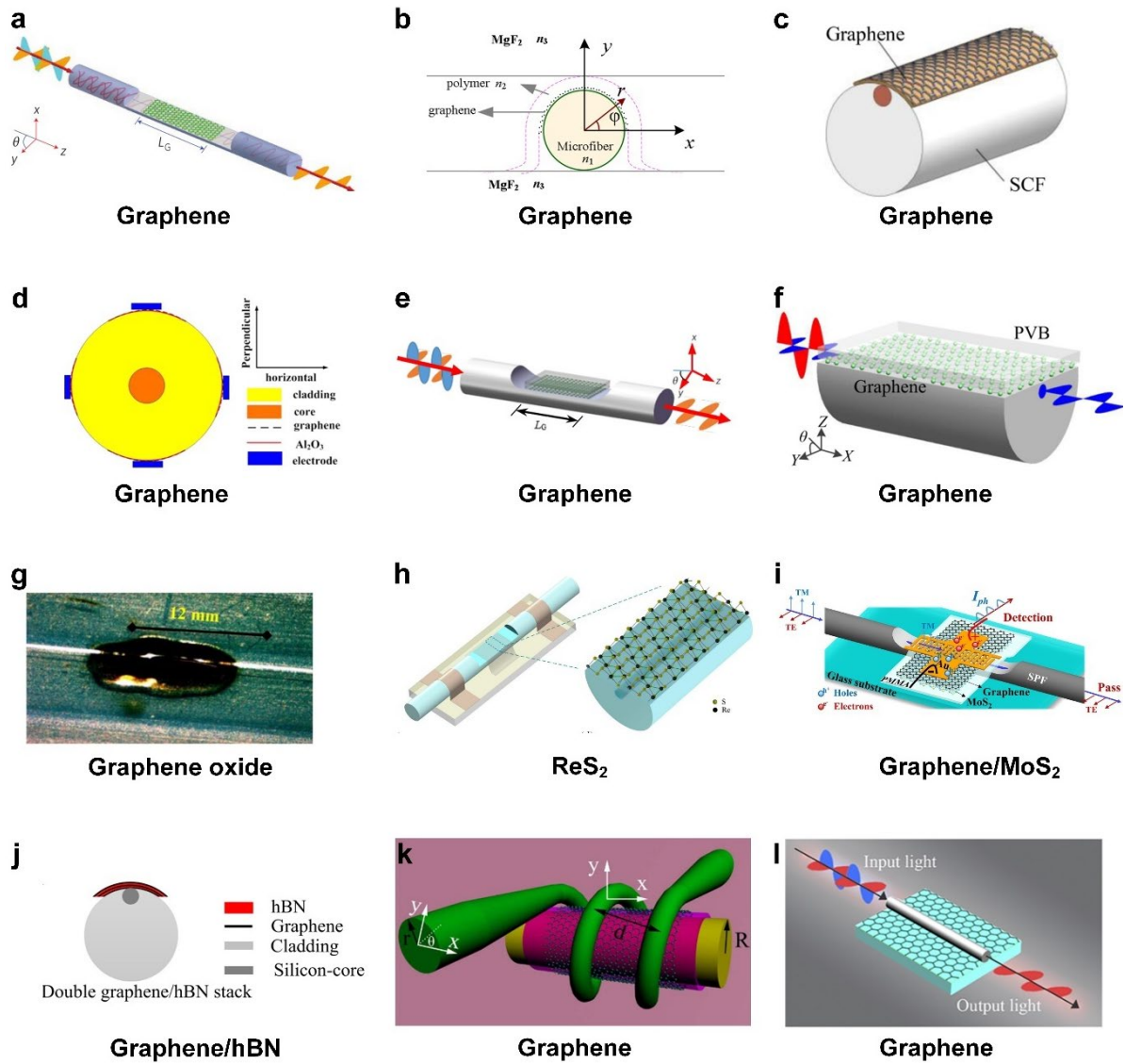

**Fig. 4. Schematic illustrations of fiber optical polarizers based on 2D materials. a.** A graphene optical polarizer with a graphene film coated on a side-polished optical fiber. **b–c.** Graphene optical fiber polarizers with engineered fiber core geometry and optimized fiber core position to enhance the light-graphene interaction. **d.** A graphene optical polarizer with graphene films coated on four sides of a microfiber surface. **e–f.** Graphene optical polarizers with auxiliar materials polymethyl methacrylate (PMMA) and polyvinyl butyral to enhance the light-graphene interaction. **g.** A graphene oxide (GO) optical polarizer with a GO film coated on a microfiber. **h.** A rhenium disulfide ($ReS_2$) optical polarizer with a $ReS_2$ film coated on a side-polished optical fiber. **i.** A vdW heterostructure optical polarizer with $MoS_2$/graphene/Au film coated on a side-polished optical fiber. **j.** A 2D heterostructure optical polarizer with graphene/hBN film coated on the surface of a silicon-core microfiber. **k.** A graphene optical polarizer with a graphene film coated on a rod wrapped by a microfiber. **l.** A graphene optical polarizer with a graphene film coated on a polydimethylsiloxane (PDMS) substrate attached with a microfiber. **a** is reprinted with permission from ref. [17] Copyright 2011 Springer Nature. **b** is reprinted with permission from ref. [95] Copyright 2014 IEEE. **c** is reprinted with permission from ref. [96] Copyright 2015 IEEE. **d** is reprinted with permission from ref. [97] Copyright 2018 IOP Publishing. **e** is reprinted with permission from ref. [98] Copyright 2017 OSA Publishing. **f** is reprinted with permission from ref.[99] Copyright 2014 IEEE. **g** is reprinted with permission from ref. [100] Copyright 2016 Taylor & Francis. **h** is reprinted with permission from ref. [101] Copyright 2017 Springer Nature. **i** is reprinted with permission from ref. [78] Copyright 2022 De Gruyter. **j** is reprinted with permission from ref.[102] Copyright 2019 IEEE. **k** is reprinted with permission from ref. [19] Copyright 2014 Optica Publishing Group. **l** is reprinted with permission from ref. [103] Copyright 2022 AIP Publishing.

### *Integrated waveguide devices*

Integrated waveguide devices with compact footprint, low power consumption, and high compatibility with photonic integrated circuits offer great benefits to achieve efficient polarization management in on-chip optical networks and photonic processing systems. Conventional metal-clad optical waveguides have been extensively used as integrated waveguide optical polarizers [104, 105]. Although they can provide high PERs, this advantage comes at the expense of a considerable IL and the demand for intricate buffer layers to facilitate a large OBW. On-chip integration of 2D materials can enhance the ability of polarizers to manipulate light with high precision and responsivity, leading to improved performance in terms of PER and OBW [18, 21, 22, 106]. **Fig. 5** shows typical integrated waveguide optical polarizers incorporating 2D materials.

Similar to spatial-light and fiber devices, graphene has been the choice for many integrated waveguide polarizers incorporating 2D materials. In 2012, a graphene waveguide polarizer was first demonstrated (**Fig. 5a**) [18]. By positioning a graphene strip supporting TE-mode surface waves on a $SiO_2$ waveguide core with an air cladding (**Fig. 5a**, left panel), the hybrid waveguide served as a TE-pass polarizer with a PER ~10 dB. By doping the graphene strip via UV-curable perfluorinated acrylate polymer resin and make it support TM-mode waves, a TM-pass polarizer was realized with a PER ~19 dB (**Fig. 5a**, right panel). In addition, a graphene/glass waveguide polarizer assisted by a ploy(methyl methacrylate) (PMMA) film was demonstrated (**Fig. 5b**) [106]. The PMMA film was used to decrease the doping level of graphene to a lower Femi level, in contrast to the use of UV-curable polymer in **Fig. 5a** to obtain a higher Femi level. By engineering the difference between the attenuation constants of the TE and TM modes in the hybrid waveguide, a high PER of ~27 dB was achieved.

In contrast the above graphene waveguide optical polarizers with the graphene layer positioned outside the waveguide cores, a graphene waveguide optical polarizer featured with a graphene film sandwiched between two chalcogenide layers was demonstrated (**Fig. 5c**) [21].

Owing to the significantly enhanced light-graphene interaction in such a device configuration, a high PER of over 20 dB extending from ~940 nm to ~1600 nm was achieved. The excess propagation losses induced by graphene for the TE and TM modes were ~590 dB/cm and ~20 dB/cm, respectively. The fabrication of the polarizer involved depositing a chalcogenide layer on a silicon wafer substrate before graphene was transferred by wet transferring method. After that, electron-beam lithography and oxygen plasma etching were employed to pattern graphene, followed by the deposition of another chalcogenide glass layer via thermal evaporation.

To simplify polarization manipulation and achieve flexible adjustment of PER in practical operation, tunable graphene waveguide optical polarizers assisted by the micro-opto-mechanical systems (MOMS) technology were demonstrated (**Fig. 5d**) [22]. In such devices, graphene layers with different thicknesses were coated on the polymer waveguide core, which were mechanically pushed to suppress the undesired polarization state. Experimental results showed that the device with a few-layer graphene film selectively attenuated TE mode and enabled the device to pass TM polarization (**Fig. 5d**, left panel), whereas the device with a thicker graphene film reduced the TM mode intensity, thus transforming the device to pass TE polarization (**Fig. 5d**, right panel). The PERs of the devices were dynamically tuned by accurate mechanical adjustment of the air gap between the graphene superstrate and the waveguide core, achieving up to ~2.5 dB and ~6.4 dB for the TE-pass and TM-pass devices, respectively.

In addition to the aforementioned experimental works, there are a lot of theoretical studies which designed graphene waveguide optical polarizers based on different structures [107-109]. A compact and broadband TM-pass polarizer was proposed based on a graphene-silicon horizontal slot waveguide structure [107]. The interaction between light and graphene was significantly enhanced by strategically placing double-layer graphene sheets in a close proximity to the slot region, allowing a theoretical PER over 40 dB in the wavelength range of 1450 – 1650 nm. Similarly, an optical polarizer based on a silicon slot waveguide with multilayer graphene inserted in the slot region was investigated [108], achieving a theoretical PER exceeding 20 dB. A tunable optical polarizer based on silicon Mach-Zehnder interferometers coated with graphene film was also proposed [109]. By varying bias voltages applied to the graphene layers, distinct alterations in the mode effective index for both TE and

TM polarizations could be achieved, allowing a theoretical PER of ~19 dB for the TE-pass state and of ~21 dB for the TM-pass state over a wide OBW ranging from ~1500 nm to ~1800 nm.

Apart from graphene, other 2D materials have also been employed in waveguide optical polarizers. A GO waveguide optical polarizer was fabricated by using drop-casting method to coat a 2-μm-thick GO film onto an SU8 polymer waveguide (**Fig. 5e**) [110], which achieved a high PER of ~40 dB over a wide wavelength range from 1530 nm to 1630 nm. In contrast to the drop-casting method that was mainly used for coating thick films (typically with a resolution of hundreds of nanometers), a layer-by-layer film coating method based on self-assembly was employed to coat GO films onto doped silica waveguides (**Fig. 5f**) [34], allowing for accurate control of the film thickness with a high resolution of ~2 nm. This, together with the use of photolithography and lift-off to precisely control the length and position of the GO films, enabled the optimization of the device performance as optical polarizers. A high PER of ~53.8 dB was achieved for a 1.5-cm-long doped silica waveguide coated with a 2-mm-long GO film. A wide OBW extending from visible (~632 nm) to infrared (~1550 nm) wavelengths was also demonstrated. By coating a 50-μm-long GO film onto a doped silica microring resonator using the same method, a microring resonator polarizer was also demonstrated (**Fig. 5g**), achieving a PER of ~8.3 dB between TE and TM-polarized resonances.

Polarization-dependent optical absorption of $MoS_2$ film on an Nd:YAG (neodymium doped yttrium aluminum garnet) waveguide was also experimentally observed (**Fig. 5h**) [111]. The $MoS_2$ film with a thickness of ~6.5 nm was coated on the waveguide via pulsed laser deposition. For a 10-mm-long waveguide uniformly coated with the $MoS_2$ film, a PER ~3 dB and a low IL ~0.4 dB was achieved. Subsequently, a $MoS_2$-coated waveguide optical polarizer was demonstrated [112], where the $MoS_2$ was coated on SU-8 polymer strip waveguide by the drop-casting method. Experimental results showed that a maximum PER of ~12.6 dB at ~980 nm was achieved. In addition, a $MoSe_2$-coated waveguide optical polarizer was demonstrated, achieving a PER of ~14 dB at ~1480 nm (**Fig. 5i**) [113]. The device was fabricated by using the drop-casting method to coat $MoSe_2$ film with a thickness of ~24 μm onto an SU8 polymer waveguide. By stacking a hybrid metamaterial structure consisting of periodic BP and silica layers on silicon substrate, a compact tunable optical polarizer was designed to operate within the visible spectrum from 500 to 800 nm (**Fig. 5j**) [114]. Simulation results showed that the polarizer was reconfigurable for either TE-pass or TM-pass by adjusting the orientation of the BP layers relative to the propagation direction.

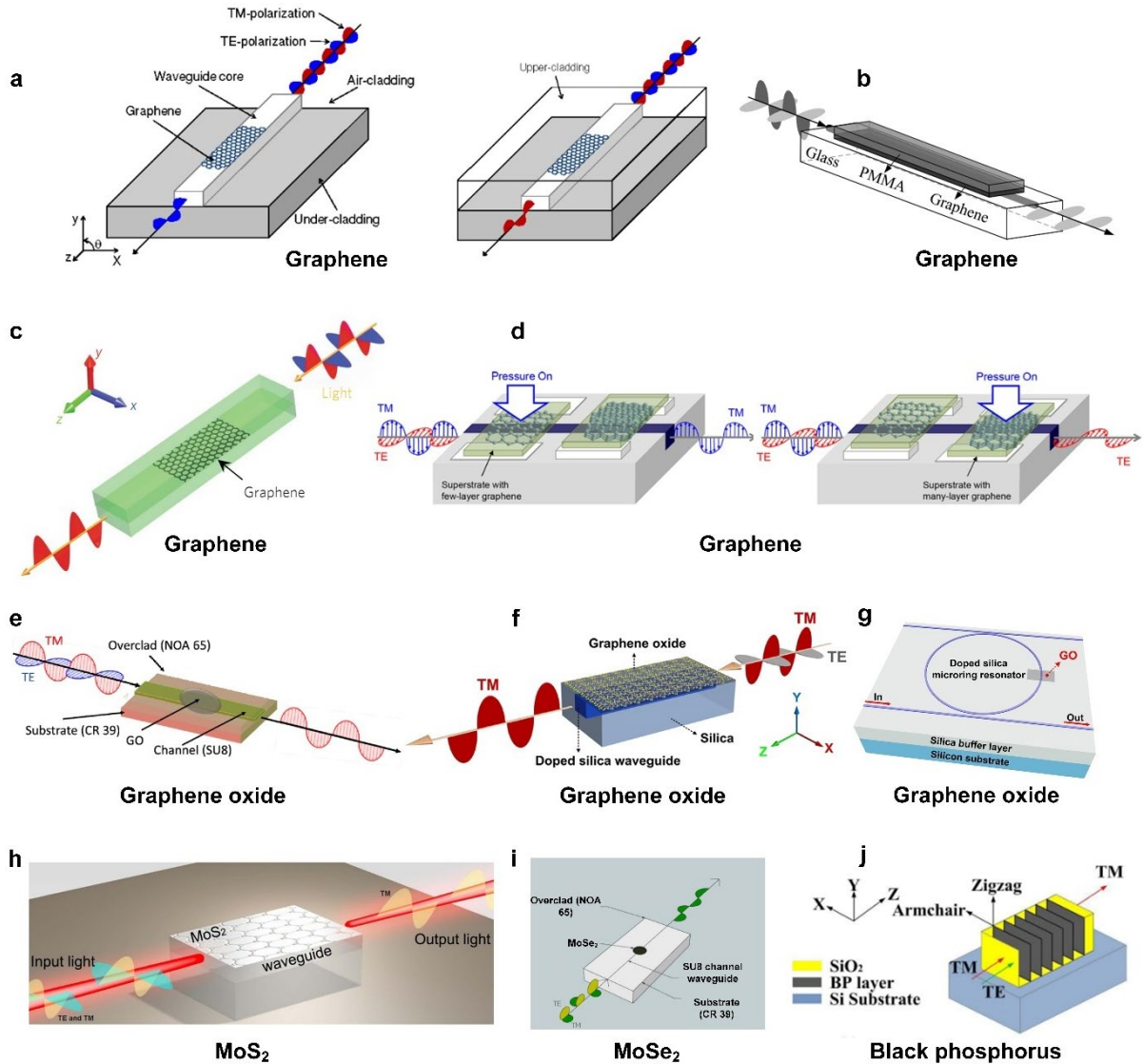

**Fig. 5. Schematic illustrations of integrated waveguide optical polarizers based on 2D materials.**
**a.** A graphene optical polarizer where a graphene strip was coated on a $SiO_2$ waveguide core with an air cladding (left panel) and with a UV-curable perfluorinated acrylate polymer resin cladding (right panel). **b.** A graphene optical polarizer with a graphene film coated on a glass waveguide assisted by a ploy(methyl methacrylate) (PMMA) film. **c.** A graphene waveguide optical polarizer with a graphene film sandwiched between two chalcogenide layers. **d.** Graphene optical polarizers assisted by the micro-opto-mechanical systems (MOMS) technology, where a few-layer graphene film (left panel) and a thicker graphene film (right panel) were coated on polymer waveguides. **e.** A graphene oxide (GO) optical polarizer with a GO film coated on an SU8 polymer waveguide. **f.** A GO optical polarizer with a GO film coated on a doped silica waveguide. **g.** A GO microring polarizer with a GO film coated on a doped silica microring resonator. **h.** A $MoS_2$ optical polarizer with a $MoS_2$ film coated on an Nd:YAG (neodymium doped yttrium aluminum garnet) waveguide. **i.** A $MoSe_2$ optical polarizer with a $MoSe_2$ film coated on an SU8 polymer waveguide. **j.** A black phosphorus (BP) optical polarizer with a periodic BP film stacking on the metamaterial waveguide. **a** is reprinted with permission from ref. [18] Copyright 2012 OSA Publishing. **b** is reprinted with permission from ref. [106] Copyright 2015 IEEE. **c** is reprinted with permission from ref. [21] Copyright 2017 Springer Nature. **d** is reprinted with permission from ref. [22] Copyright 2018 Wiley-VCH Verlag GmbH & Co. **e** is reprinted with permission from ref. [110] Copyright 2014 OSA Publishing. **f-g** are reprinted with permission from ref. [34] Copyright 2019 Wiley-VCH Verlag GmbH & Co. **h** is reprinted with permission from ref. [111] Copyright 2014 Springer Nature. **i** is reprinted with permission from ref. [113] Copyright 2019 Elsevier B.V.. **j** is reprinted with permission from ref. [114] Copyright 2017 IEEE.

In **Table 1**, we summarize 2D-material-based optical polarizers in different device platforms and compare their performance. Here we compare the key performance parameters including PER, OBW, and IL, and only show the results for experimental works. From **Table 1**, it can be seen that the spatial-light and fiber optical polarizers require relatively thick films of 2D materials as compared to the integrated waveguide polarizers. This is mainly due to the relatively weak mode overlap with 2D materials in these devices. It is also worth noting that even for some devices with very thick films beyond the typical thickness of 2D materials, the anisotropy in the film loss can still remain sufficiently high to facilitate their functions as optical polarizers.

In contrast to fiber and integrated waveguide polarizers that confine light within waveguides with mode dispersion, spatial-light optical polarizers have the ability to operate in free space that has almost no mode dispersion. In addition, they can be designed to function across different wavelength regions and typically demonstrate wider OBWs compared to fibers and integrated waveguide devices. On the other hand, the strong mode confinement in fiber and integrated waveguide optical polarizers makes them attractive in achieving compact device footprint, and the relatively strong interaction between the light waves and the 2D materials in these devices also yield relatively high PER values compared to the spatial-light devices.

Regarding the 2D materials selected for these optical polarizers, most studies opted for graphene, likely due to its earlier discovery and more extensive investigation. Along with the rapid advancements in the studies and fabrication of 2D materials over the past decade, an increasing variety of 2D materials are being utilized for implementing optical polarizers, which demonstrate their own advantages and address a broader application scope. The future development in this field hinges on overcoming the specific limitations to materials and platforms, thereby paving the way for more versatile and broadly applicable polarizing technologies.

**Table 1 Comparison of 2D-material-based polarizers in different device platforms.** PER: polarization extinction ratio. OBW: operational bandwidth. IL: insertion loss. WG: waveguide.

| Materials | Platforms | Thickness | PER (dB) | OBW (μm) | IL (dB) | Ref. |
|---|---|---|---|---|---|---|
| Graphene | Spatial-light | Monolayer | – [a] | ~188 – 500 | – [a] | [20] |
| Graphene | Spatial-light | Monolayer | ~10 | ~120 – 600 | – [a] | [26] |
| BP | Spatial-light | ~96 nm | ~9 | – [a] | – [a] | [90] |
| GO | Spatial-light | ~1000 nm | ~20 | ~2 – 14 | – [a] | [89] |
| MXene | Spatial-light | ~30 nm | ~6 | ~150 – 1000 | <2.0 | [66] |
| Perovskite | Spatial-light | ~40 nm | – [a] | ~0.20 – 0.60 | – [a] | [75] |
| Graphene | Fiber | Monolayer | ~27 | ~0.50 – 1.63 | ~5.0 | [17] |
| Graphene | Fiber | ~1.0 nm | ~16 | ~1.20 – 1.65 | – [a] | [19] |
| Graphene | Fiber | Monolayer | ~38 | ~1.43 – 2.00 | ~1.0 | [99] |
| Graphene | Fiber | 2 layers | ~44 | ~1.56 – 1.63 | >5.0 | [98] |
| Graphene | Fiber | Monolayer | ~31 | – [a] | – [a] | [103] |
| Graphene/$MoS_2$ | Fiber | Monolayer | ~19 | ~0.98 – 1.62 | <10.5 | [78] |
| GO | Fiber | ~24000 nm | ~36 | ~1.30 – 1.60 | ~2.2 | [100] |
| $ReS_2$ | Fiber | Monolayer | – [a] | – [a] | <4.4 | [101] |
| Graphene | Polymer WG | – [a] | ~19 | – [a] | ~26.0 | [18] |
| Graphene | Glass WG | – [a] | ~27 | ~1.23 – 1.61 | ~9.0 | [106] |
| Graphene | Chalcogenide glass-on-graphene WG | Monolayer | ~23 | ~0.94 – 1.60 | ~0.8 | [21] |
| Graphene | Polymer WG | > or < 10 nm [b] | ~6 | – [a] | ~9.0 | [22] |
| GO | Polymer WG | ~2000 nm | ~40 | ~1.53 – 1.63 | ~6.5 | [110] |
| GO | Doped silica WG | ~2 – 200 nm | ~54 | ~0.63 – 1.60 | ~7.5 | [34] |
| $MoS_2$ | Nd:YAG waveguide | ~6.5 nm | ~3 | – [a] | ~0.4 | [111] |
| $MoS_2$ | Polymer WG | ~2.5 nm | ~12.6 | ~0.65 – 0.98 | <10 | [112] |
| BP | Metamaterial WG | ~0.5 nm | ~30 | ~0.50 – 0.80 | ~0.3 | [114] |
| $MoSe_2$ | Polymer WG | ~24000 nm | ~14 | ~0.98 – 1.55 | – [a] | [113] |

60. [a] There is no reported value for this parameter in the literature.
61. [b] The polymer waveguides with a few-layer graphene film (< 10 nm) and a thicker graphene film (> 10 nm) worked as TM- and TE-pass optical polarizers, respectively.

## Challenges and perspectives

As can be seen from the substantial works reviewed in previous section, the past decade has seen remarkable advancements in optical polarizers based on 2D materials. Despite the significant progress, this field still faces challenges and has new demands for further developments. In this section, we discuss the current challenges and future opportunities.

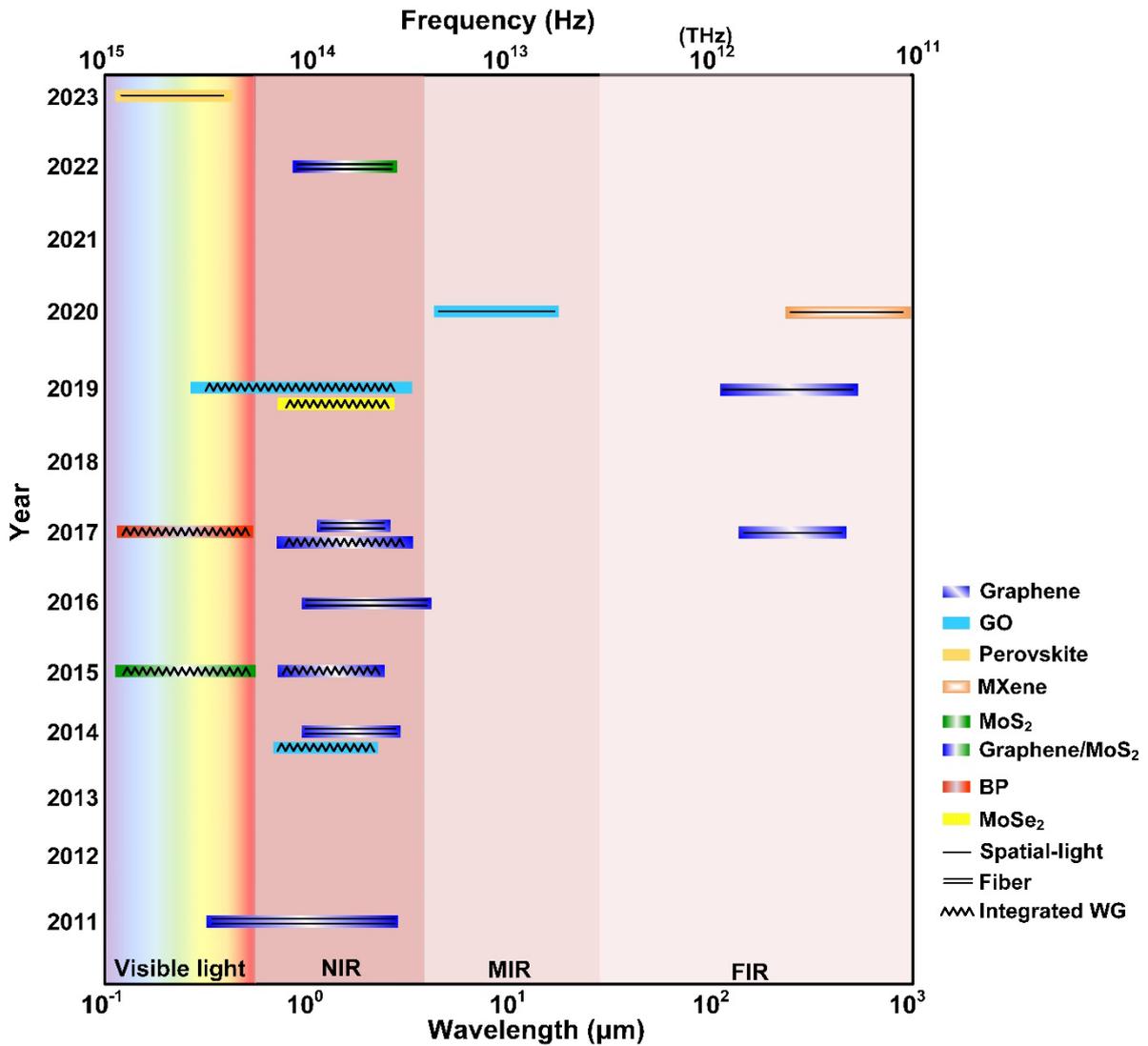

Fig. 6. Development roadmap of 2D-material-based optical polarizers with different platforms. WG: waveguide. Values of wavelength ranges are taken from: Ref. [17] for 2011 (Graphene), Ref. [110] for 2014 (GO), Ref. [19] for 2014 (Graphene), Ref. [112] for 2015 (MoS$_2$), Ref. [106] for 2015 (Graphene), Ref. [99] for 2016 (Graphene), Ref. [114] for 2017 (BP), Ref. [21] for 2017 (Graphene, integrated WG), Ref. [98] for 2017 (Graphene, fiber), Ref. [20] for 2017 (Graphene, spatial-light), Ref. [34] for 2019 (GO), Ref. [113] for 2019 (MoSe$_2$), Ref. [26] for 2019 (Graphene), Ref. [89] for 2020 (GO), Ref. [66] for 2020 (MXene), Ref. [78] for 2022 (Graphene / MoS$_2$), Ref.[75] for 2023

(Perovskite).

**Fig. 6** shows a development roadmap of optical polarizers based on 2D materials, which includes typical experimental works on different 2D materials and in different device platforms. Here, we focus on comparing the operating wavelength ranges of these devices, and it can be seen that the fiber and integrated waveguide optical polarizers are limited to work within the visible and near-infrared regions. This is mainly because these devices with compact footprint and small mode area cannot support the propagation of THz waves. In contrast, spatial-light devices dominate the optical polarizers working in the THz region. In these devices with light propagating freely in space, more efficient interaction with THz waves can be achieved. Developing THz optical polarizers with a more compact footprint is a direction for future advancements. To achieve this, incorporating 2D materials into plasmonic devices that provide sub-wavelength confinement beyond the diffraction limit [115, 116] could be a possible approach, while the yield needs to be balanced with the relatively high propagation loss in these devices.

In practical applications, the stability of 2D materials is critical for devices incorporating them, and this is also true for 2D-material-based optical polarizers. 2D materials with large surface area and ultrathin film thicknesses may be susceptible to environmental factors including changes in temperature, humidity, mechanical strain, and exposure to chemicals. This sensitivity is particularly pronounced in 2D materials like BP and TMDCs, which exhibit limited stability in air. To improve the environmental stability of 2D materials, two strategies are usually employed. The first is adopting encapsulation materials to effectively isolate the 2D materials from ambient conditions. For instance, dielectrics (*e.g.*, $Al_2O_3$) [117] and polymers (*e.g.*, PMMA and PVB) [118, 119] were used as encapsulation layers. The second involves introducing robust 2D materials such as hBN and GO to render the surface of sensitive 2D materials inert to environmental influences [120]. For example, few-layer holey GO membranes were utilized as passivation layers to protect BP and MXene from oxidative degradation [121]. It should also be noted that the incorporation of polymer or other 2D materials might decrease thermal dissipation in layered 2D films, which should be considered for devices operating at high light intensities and with temperature-sensitive 2D materials.

The trade-off between PER and IL presents another challenge in the design and fabrication of 2D-material-based optical polarizers, particularly for fiber and integrated waveguide devices. Typically, a device with a thicker 2D material film exhibits a higher PER and IL, where the former is advantageous, but the latter is undesirable. In some literatures, the figure of merit (FOM) for optical polarizers based on 2D materials was defined as the ratio of PER to IL [21, 34]. To balance the trade-off between these two parameters, it is important to engineer the mode overlap with 2D materials, and this can be achieved by optimizing the waveguide geometry as well as the thickness and position of the 2D films. Given that the IL accumulates over a certain length, it can always be minimized by shortening the lengths of the 2D films, although this may also result in decreased PERs. Recently, some methods to manipulate the optical anisotropy of 2D materials have been proposed [13, 122]. For example, by adjusting the twist angle between adjacent layers, BP exhibited distinctive properties in its optical anisotropy [123]. The change of the intrinsic optical anisotropy of 2D materials can also be realized by introducing external factors, such as stain, electric field, and surface plasmon [124-126].

In terms of device fabrication, currently there are mainly two fabrication strategies to incorporate 2D materials into different device platforms. The first involves the transfer of 2D materials to the target device substrates, which is represented by the dry transfer method (using a transfer stamp [127, 128] to obtain 2D films from bulk materials) and the wet transfer method (typically for CVD-grown 2D materials on metal substrates or foils [129, 130]). The second features directly coating 2D materials onto the target devices, which is exemplified by the solution dropping method [131] and the self-assembly method [38]. Despite the wide use of the transfer methods in lab experiments, they face limitations in achieving high efficiency for large-scale manufacturing and precise control of the 2D film parameters. In addition, the transfer process can easily lead to stretching, bending, and wrinkling of the 2D films, posing challenges in achieving high film uniformity. On the other hand, although the solution dropping method offer a simple and rapid way to coated 2D material films onto the target devices, it faces limitations due to low film uniformity and considerable film thickness. In contrast, the self-assembly coating method allows for layer-by-layer film coating with precise control over the thicknesses of 2D material films. It also shows attractive abilities to achieve large-area film coating, high film uniformity, and conformal coating on complex structures. To date, the self-assembly method has been employed for coating graphene, GO, TMDC, and MXene films, with more 2D materials remaining to be explored [34, 39, 132, 133].

The precise control of 2D material film parameters, such as thickness, length, and placement, is crucial for optimizing the performance of optical polarizers. The ability to pattern 2D materials can also introduce more complex polarization functionalities that go beyond what traditional polarization mechanisms can provide. For example, uniformly spaced and well-aligned free-standing 2D materials patterns are needed for 2D-material-based wire grid polarizers to achieve a high PER [75, 88], and a THz spatial light optical polarizer with graphene patches in different patterns can convert the linearly polarized waves into different polarization states [87]. So far, many approaches have been employed to pattern 2D materials, such as laser patterning, lithography, nanoimprinting, inkjet printing, and focused ion beam (FIB) milling [134-137]. Each of these approaches has its advantages for specific purposes. For example, the laser patterning approach provides a mask-free and chemical-free method with simple fabrication process, along with the ability to pattern a variety of 2D materials. The lithography and nanoimprinting approaches leverage the well-developed patterning techniques for the fabrication of integrated circuits, and show better compatibility with the integrated chip industry. Inkjet printing allows for rapid and in situ patterning, and is attractive for fabricating large-area patterns with relatively low resolution (typically > 1 μm [138]). FIB milling can achieve ultrahigh patterning resolution (< 10 nm [139]), but this normally comes at the expense of a lower patterning speed than other patterning approaches.

Although in this review we focus on optical polarizers based on 2D materials, it is worth noting that 2D materials have also been employed to improve the performance of other polarization sensitive devices such as photodetectors, light-emitting devices, and mode-locked lasers. **Fig. 7** shows typical polarization sensitive devices incorporating 2D materials. Polarization-sensitive photodetectors with 2D materials such as perovskite, germanium selenide ($GeSe_2$), germanium arsenide ($GeAs_2$), and MXene have been reported **(Fig. 7a)** [65, 140-142]. The 2D materials with strong optical anisotropy in these photodetectors allow them to exhibit prominent ability to capture light with high responsivity, fast response speed, and broad response bandwidth. In addition, 2D materials such as TMDCs and BP have been utilized in polarization sensitive light emitting devices (**Fig. 7b**) [143-146]. The direct bandgap of these 2D materials allows for photon emission upon excitation, and their optical anisotropy offers the capability to control the polarization state of the emitted light. 2D materials such as BP and GO have been employed in mode-locked fiber lasers (**Fig. 7c**) [147, 148], where the 2D materials featured with both high saturable absorption and high anisotropic optical absorption enable the generation of mode-locked optical pulses via simple polarization control. In addition to the above devices, 2D materials such as graphene, BP, and vdW heterostructures have also been used for implementing other polarization sensitive optical devices, such as converters, waveplates, beam splitters, and polarization switches (**Figs. 7d-7g**) [149-152]. These results have wide applications to GO based devices as well as other novel photonic platforms. [153-200] Ultimately this could be useful for both classical and quantum microcomb based applications. [201-277]

Finally, while 2D materials have already seen extensive uses in optical polarizers and polarization sensitive devices, the quest for enhanced performance of these devices continues. To further improve performance, there are mainly two strategies. One involves modification of the properties of 2D materials (*e.g.*, by optimizing their fabrication processes) to enhance their optical anisotropy, and the other is to optimize the device configuration and facilitate more efficient interaction between light and the anisotropic 2D materials. In addition, with the rapid expansion of the 2D materials family, there will also be new opportunities for the development of optical polarizers incorporating newly emerging 2D materials. Considering the extensive applications of optical polarizers highlighted at the start of this review, the continuous advancement of optical polarizers incorporating 2D materials is expected to significantly impact and facilitate a broad range of optical systems with miniatured size and reduced power consumption.

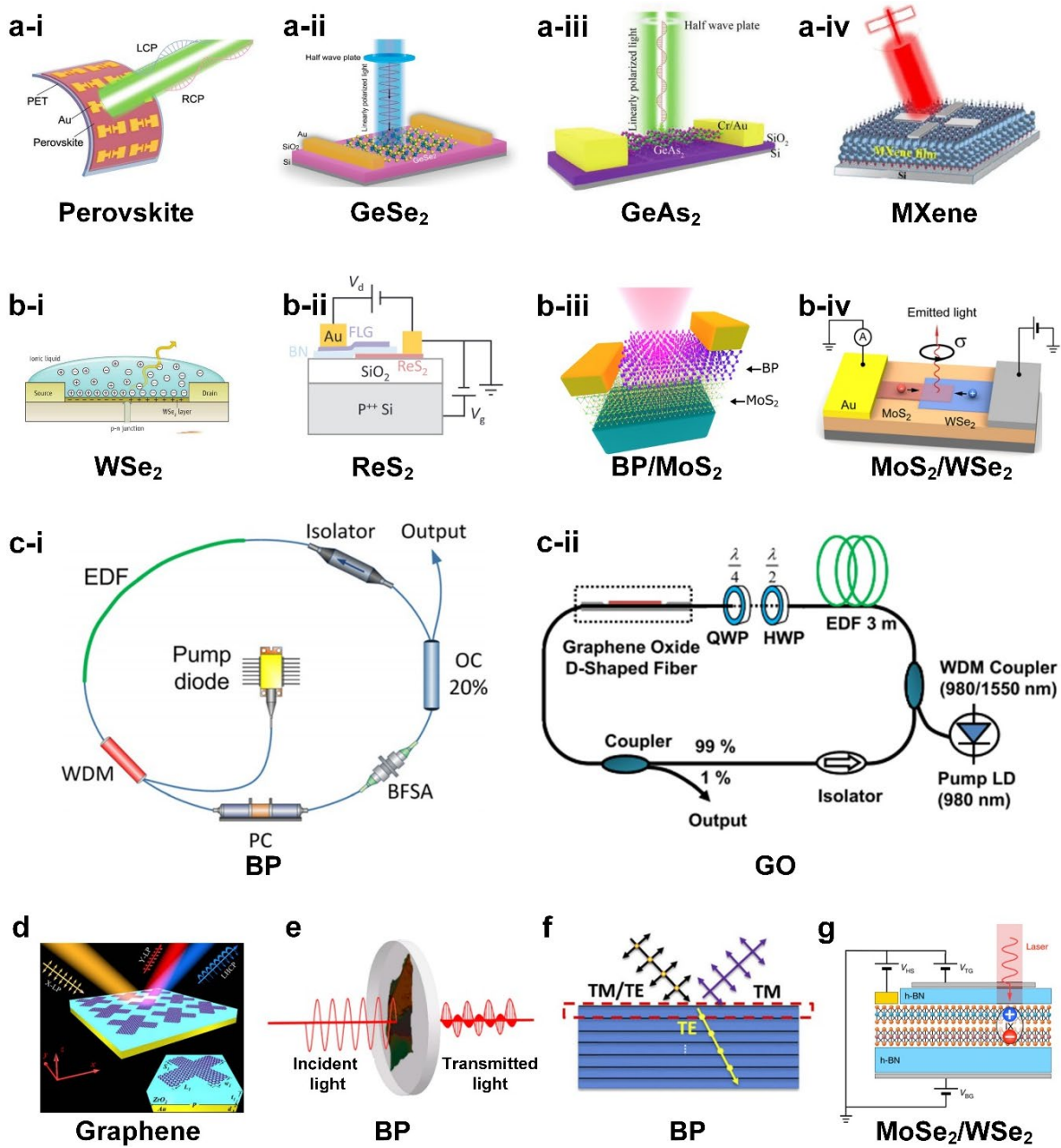

Fig. 7. Typical polarization-sensitive devices incorporating 2D materials: **a.** Photodetectors with (i) perovskite, (ii) germanium selenide (GeSe$_2$), (iii) germanium arsenide (GeAs$_2$), and (iv) MXene **b.** Light emitting devices with (i) tungsten diselenide (WSe$_2$), (ii) rhenium disulfide (ReS$_2$), (iii) black phosphorus (BP) / molybdenum disulfide (MoS$_2$), and (iv) MoS$_2$ / WSe$_2$. **c.** Mode-locked fiber lasers with (i) BP and (ii) graphene oxide (GO). **d.** Converters with graphene. **e.** Waveplates with BP. **f.** Beam splitters with BP. **g.** Polarization switches with molybdenum diselenide (MoSe$_2$)/ WSe$_2$. **a-i** is reprinted with permission from ref. [140] Copyright 2020 Wiley-VCH Verlag GmbH & Co. **a-ii** is reprinted with permission from ref. [141] Copyright 2018 American Chemical Society. **a-iii** is reprinted with permission from ref. [142] Copyright 2018 Wiley-VCH Verlag GmbH & Co. **a-iv** is reprinted with permission from ref. [65] Copyright 2022 Springer Nature. **b-i** is reprinted with permission from ref. [143] Copyright 2014 AAAS. **b-ii** is reprinted with permission from ref. [144] Copyright 2020 Wiley-VCH Verlag GmbH & Co.. **b-iii** is reprinted with permission from ref. [145] Copyright 2016 American Chemical Society. **b-iv** is reprinted with permission from ref. [146] Copyright 2016 American

Chemical Society. **c-i** is reprinted with permission from ref. [147] Copyright 2015 AIP Publishing. **c-ii** is reprinted with permission from ref. [148] Copyright 2013 IOP Publishing. **d** is reprinted with permission from ref. [149] Copyright 2021 OSA Publishing. **e** is reprinted with permission from ref. [150] Copyright 2017 American Chemical Society. **f** is reprinted with permission from ref. [151] Copyright 2020 OSA Publishing. **g** is reprinted with permission from ref. [152] Copyright 2019 Springer Nature.

## Conclusion

2D-material-based optical polarizers represent an emerging interdisciplinary field at the intersection of polarization optics and material science. Over the past decade, significant advancements in this field have been made, facilitated by the distinctive properties of advanced 2D materials and the rapid progress in their fabrication technologies. In this review, we provide an overview for recent progress in 2D-material-based optical polarizers. First, we introduce the distinctive material properties of 2D materials for implementing optical polarizers. Next, we review 2D-material-based optical polarizers in different device platforms, including spatial-light devices, fiber devices, and integrated waveguide devices. Finally, we discuss the current challenges and future perspectives of this field. In the future, along with advances in device fabrication technologies and expanded applications with demanding requirements, we believe that there will be more and more new breakthroughs in this field.

## Competing interests

The authors declare no competing interests.